\documentclass[conference]{IEEEtran}
\IEEEoverridecommandlockouts
\usepackage{cite}
\usepackage{amsmath,amssymb,amsfonts}
\usepackage{algorithmic}
\usepackage{graphicx}
\usepackage{textcomp}
\usepackage{xcolor}
\usepackage{caption}
\usepackage{subcaption}

\usepackage{nameref,cleveref}
\usepackage{mathtools}

\Crefname{thm}{Theorem}{Theorems}
\Crefname{prop}{Proposition}{Propositions}
\Crefname{lem}{Lemma}{Lemmas}
\Crefname{cor}{Corollary}{Corollaries}
\Crefname{defn}{Definition}{Definitions}
\Crefname{assump}{Assumption}{Assumptions}
\Crefname{conj}{Conjecture}{Conjectures}

\Crefname{equation}{}{}

\Crefname{figure}{Fig.}{Figs.}
\creflabelformat{figure}{#2\textup{#1}#3}

\def\BibTeX{{\rm B\kern-.05em{\sc i\kern-.025em b}\kern-.08em
    T\kern-.1667em\lower.7ex\hbox{E}\kern-.125emX}}
\begin{document}

\title{Application of Log-Linear Dynamic Inversion Control to a Multi-rotor}

\author{Li-Yu Lin, James Goppert, and Inseok Hwang
\thanks{This work is supported in part by NASA University Leadership Initiative Project on Secure and Safe Assured Autonomy (S2A2) under Grant 80NSSC20M0161.}
\thanks{L. Lin, J. Goppert, and I. Hwang are with the School of Aeronautics and Astronautics, Purdue University, West Lafayette, IN 47906, USA (e-mail: lin1191@purdue.edu; jgoppert@purdue.edu; ihwang@purdue.edu).}}

\maketitle

\begin{abstract}
This paper presents an approach that employs log-linearization in Lie group theory and the Newton-Euler equations to derive exact linear error dynamics for a multi-rotor model, and applies this model with a novel log-linear dynamic inversion controller to simplify the nonlinear distortion and enhance the robustness of the log-linearized system. In addition, we utilize Linear Matrix Inequalities (LMIs) to bound the tracking error for the log-linearization in the presence of bounded disturbance input and use the exponential map to compute the invariant set of the nonlinear system in the Lie group. We demonstrate the effectiveness of our method via an illustrative example of a multi-rotor system with a reference trajectory, and the result validates the safety guarantees of the tracking error in the presence of bounded disturbance.
\end{abstract}


\section{Introduction}
% UAM -> multi-rotor -> safety verification -> invariant set
Multi-rotors have played an important role in the Urban Air Mobility (UAM) system and have been deployed in various applications, including package delivery and aerial inspections. However, the safety of these vehicles is critical, and methods for safety verification must be developed to ensure public trust and acceptance of UAM. One promising approach is to utilize the invariant set for the dynamical system~\cite{blanchini1999set,khalil2002nonlinear,coogan2014dissipativity}, which represents a set that the system will never leave if the initial conditions are within the set. This technique can be highly effective in ensuring the safety of multi-rotor systems, making them more reliable and trustworthy for a range of applications.

% lyapunov
% linear time-varying backstepping exists, but we don't actually know the function of time in our system,
Lyapunov functions can be used to construct invariant sets for dynamical systems~\cite{khalil2002nonlinear}. While nonlinear systems can be analyzed using hand-crafted Lyapunov functions or Lyapunov-based control design methods like backstepping control~\cite{wang2016backstepping}, these approaches are challenging to generalize and can be tedious to implement. For linear systems with bounded input, however, invariant sets can be constructed using backstepping control or Linear Matrix Inequality (LMI) techniques~\cite{boyd1994linear}. Although linear time-varying backstepping exists, it requires the disturbance to be a known function of time; however, the function of time for the disturbance is unknown in the multi-rotor system, the linear time-varying backstepping cannot be used in our case. Since the use of LMI is a more efficient method for finding the Lyapunov function and constructing the invariant set, we propose utilizing the linearization of the multi-rotor system to apply LMI techniques and construct the invariant set.

% lie group
The general linearization technique cannot give an exact solution, and thus the invariant set obtained must be over-approximated to account for linearization error. To address this issue, we employ the log-linearization technique from Lie group theory~\cite{barrau2016invariant, barfoot2014associating}. While previous studies~\cite{barrau2016invariant, teng2022lie} have utilized the log-linear property for linearization, this technique only provides an approximation, not an exact solution. In our previous work~\cite{lin2022correct}, we proposed a generalized method for exact linearization of the nonlinear kinematics model in the Lie group to a corresponding linear system in the Lie algebra, using the derivative of the exponential map~\cite{rossmann2006lie}. Since the result is an exact linearization, it allows us to analyze the nonlinear dynamics in a more efficient way using the linear dynamics in the Lie algebra, which is a vector space. A similar work in~\cite{li2022closed}, derives the exact log-linearization for a specific Lie group, $SE_2(3)$, our approach in~\cite{lin2022correct} applies to all matrix Lie groups. Since the kinematic model alone is insufficient for describing the behavior of a multi-rotor system, we employ the $SE_2(3)$ Lie group~\cite{barrau2016invariant, luo2020geometry}, which is a group of double homogeneous matrices, as part of our model.

This paper presents an efficient method to compute the invariant set for a nonlinear multi-rotor system with feedback control and subject to disturbance. Our method is based on log-linearity in the $SE_2(3)$ Lie group and LMI theory. To achieve this, we first embed the multi-rotor model in the $SE_2(3)$ Lie group and employ log-linearization to the left-invariant error dynamics. However, since the differential equation governing the evolution of the angular velocity derived from the Newton-Euler equations cannot be embedded in the $SE_2(3)$ Lie group, we treat this as a separate sub-system. Moreover, to further enhance the robustness and simplify the system, we follow our previous work~\cite{lin2022correct} to design a log-linear dynamic inversion based control law for a multi-rotor with provable safety guarantees in the presence of bounded disturbance. Previous work in \cite{sola2018micro} shows a succinct synopsis of Lie group theory as it applies to state estimation and in addition, also the control methods we use in this paper.

The rest of this paper is organized as follows. In \Cref{sec:II}, we embed the three-dimensional multi-rotor model in the $SE_2(3)$ Lie group. In \Cref{sec:III}, we derive the error dynamics of log-linearization and the angular velocity, and design the dynamic inversion control for both systems. In \Cref{sec:IV}, we illustrate the simulation results of the invariant set for our model with two different scenarios. Finally, in \Cref{sec:V}, we present our concluding remarks and immediate future works.

\section{Vehicle Dynamics Embedded in $SE_2(3)$ Lie group}
\label{sec:II}
Consider the multi-rotor model evolving in a 3D space given as:
\begin{equation}
    \frac{d}{dt}p = v,\;\frac{d}{dt}v = g + R a,\;\frac{d}{dt}R = R\omega_\times,\;\frac{d}{dt}\omega = \alpha
\label{eq:sys}
\end{equation}
where $R \in \mathbb{R}^{3\times3}$ and $p \in \mathbb{R}^{3\times1}$ denote the vehicle attitude and position, $\omega \in \mathbb{R}^{3\times1}$ and $v \in \mathbb{R}^{3\times1}$ denote the angular and translational velocity, $g \in \mathbb{R}^{3\times1}$ denotes the gravity, $a \in \mathbb{R}^{3\times1}$ and $\alpha \in \mathbb{R}^{3\times1}$ denote the angular and translational acceleration respectively. $\omega_\times$ represents the skew symmetric matrix of $\omega$.

From Barrau's paper\cite{barrau2016invariant}, we can embed the attitude, position, and translational velocity to a group of double homogeneous matrices $SE_2(3)$. The dynamics can be written in the mixed-invariant system~\cite{khosravian2016state, lin2022correct}, as follow:
\begin{equation}
    \dot{X} = X (C+[\nu_v]^\wedge) + ([\nu_g]^\wedge - C) X
\label{eq:dX}
\end{equation}
where $X$ represents the state of the 3D vehicle in the $SE_2(3)$ Lie group. Note that $\nu_v = \begin{bmatrix} 0 & a & \omega\end{bmatrix}^T$ and $\nu_g = \begin{bmatrix} 0 & g & 0\end{bmatrix}^T$ may be functions of time and represent the vehicle input, including feedback control and disturbance, and $g$ denotes gravity. $C = \begin{bmatrix} 0_{3\times3} & 0_{3\times1} & 0_{3\times1}\\ 0_{1\times3} & 0 & 1 \\  0_{1\times3} & 0 & 0 \end{bmatrix}$ is a constant matrix, which embeds the kinematic equation $\frac{d}{dt}p = v$.

The $SE_2(3)$ Lie group can be represented by a matrix of the form:
\begin{equation}
    X = \begin{bmatrix}
        R & v & p \\
        0 & 1 & 0 \\
        0 & 0 & 1
        \end{bmatrix}
\end{equation}
The corresponding $\mathfrak{se}_2(3)$ Lie algebra can be represented by a matrix of the form:
\begin{equation}
    [x]^\wedge = \begin{bmatrix}
                 \omega_\times & a & v \\
                 0 & 0 & 0 \\
                 0 & 0 & 0 
                 \end{bmatrix}   
\end{equation}
where $x = \begin{bmatrix}v & a & \omega\end{bmatrix}^T$ is the element in the Lie algebra, and $[\cdot]^\wedge$ indicates the wedge operator that maps the element from $\mathbb{R}^9$ to the $\mathfrak{se}_2(3)$ Lie algebra. Although the angular velocity, $\omega$, cannot be embedded in the $SE_2(3)$ Lie group, we can consider the dynamics of the angular velocity as a separate sub-system in our application.

\section{Control Design for Error Dynamics}
\label{sec:III}
In this section, we first derive the log-linear property for the error dynamics in the $SE_2(3)$ Lie group. Based on the log-linear property, we then log-linearize the nonlinear system with feedback control, disturbance, and noise to a linear bounded input system. We then derive the error dynamics for the angular velocity, $\omega$, with different frame considerations for the vehicle and reference systems.

\subsection{Log-Linearization of Error Dynamics in the $SE_2(3)$ Lie Group}
Consider the dynamics of two systems, where $X_b \in SE_2(3)$ is the state of the vehicle system in the vehicle body frame, which is denoted by $_b$, and $\bar{X}_r \in SE_2(3)$ is the state of the reference trajectory in the reference frame, which is denoted by $_r$:
\begin{align}
    \dot{X_b} &= X_b (C+[\nu_{v_b}]^\wedge) + ([\nu_{g_e}]^\wedge - C) X_b \\
    \dot{\bar{X}}_r &= \bar{X}_r (C+[\bar{\nu}_{v_r}]^\wedge) + ([\nu_{g_e}]^\wedge - C) \bar{X}_r
\end{align}
where $\nu_{v_b}, \; \bar{\nu}_{v_r}\in \mathfrak{se}_2(3)$ are system inputs and may be functions of time. The relationship between the vehicle and reference input, $\nu_{v_b}$ and $\bar{\nu}_{v_r}$, is $\nu_{v_b} = \bar{\nu}_{v_r} + \tilde{\nu}$, where $\tilde{\nu}$ represents the difference between two inputs, which includes feedback control inputs, and disturbances. $\nu_{g_e} \in \mathfrak{se}_2(3)$ represents the gravity in the earth frame. Applying the chain rule in the left-invariant error, $\eta = X_b^{-1}\bar{X}_r$, the dynamics of the left-invariant error can be written as:
\begin{equation}
    \dot{\eta} = \eta (C+[\bar{\nu}_{v_r}]^\wedge) - (C+[\nu_{v_b}]^\wedge) \eta
\end{equation}

Based on our previous work~\cite{lin2022correct}, we denote the left-invariant error $\eta$ by $\eta = \exp{[\zeta]^\wedge}$. Applying the derivative of the exponential map~\cite{rossmann2006lie}, we can observe the error dynamics in the $\mathfrak{se}_2(3)$ Lie algebra, as follows:
\begin{equation}
\begin{aligned}
    [\dot{\zeta}]^\wedge &= ad_{[\zeta]^\wedge} (C+[\bar{\nu}_{v_r}]^\wedge) +  [U_\zeta \tilde{\nu}]^\wedge \\
    U_\zeta &\equiv -\frac{ad_{[\zeta]^\wedge} \exp{(-ad_{[\zeta]^\wedge})}}{I - \exp{(-ad_{[\zeta]^\wedge})}}
\end{aligned}   
\end{equation}
where $\zeta = \begin{bmatrix}
    \zeta_p & \zeta_v & \zeta_R
\end{bmatrix}^T$ is the state of error dynamics in the $\mathfrak{se}_2(3)$ Lie algebra, and $U_\zeta$ is the matrix of nonlinear distortion of inputs in the $\mathfrak{se}_2(3)$ Lie algebra for the left-invariant error dynamics.

The adjoint function of $\zeta$ operating on $C$ can be written as a linear operator of $C^\triangle$ operating on $\zeta$:
\begin{align}
    ad_{[\zeta]^\wedge} C &= -[C^\triangle \zeta]^ \wedge \\
    C^\triangle &= \begin{bmatrix}
        0_{3\times3} & I_{3} & 0_{3\times3} \\
        0_{3\times3} & 0_{3\times3} & 0_{3\times3} \\
        0_{3\times3} & 0_{3\times3} & 0_{3\times3}
    \end{bmatrix}
\end{align}
and the adjoint linear operator of an $\mathfrak{se}_2(3)$ Lie algebra element, $\bar{\nu}_{v_r}$, can be written as:
\begin{equation}
    ad_{[\bar{\nu}_{v_r}]^\wedge} = \begin{bmatrix}
        \bar{\omega}_\times & 0_{3\times3} & \bar{v}_\times \\
        0_{3\times3} & \bar{\omega}_\times & \bar{a}_\times \\
        0_{3\times3} & 0_{3\times3} & \bar{\omega}_\times
    \end{bmatrix}
\end{equation}

We apply the Lie vee operator, $[\cdot]^\vee$, an inverse of the Lie wedge operator, to map the error dynamics from a Lie algebra matrix representation to elements of the vector space $\mathbb{R}^9$, as follows:
\begin{equation}
\begin{aligned}
\dot{\zeta} &= -(ad_{[\bar{\nu}_{v_r}]^\wedge} + C^\triangle) \zeta +  U_\zeta \tilde{\nu}
\end{aligned}
\end{equation}

Here, we want to control the angular velocity and acceleration in the thrust, which correspond to the moment and force for the multi-rotor; therefore, the system can be written as:
\begin{equation}
\begin{aligned}
    \dot{\zeta} &= -(ad_{[\bar{\nu}_{v_r}]^\wedge} + C^\triangle) \zeta +  B_u U_\zeta u_\zeta + B_d U_\zeta d_\nu \\
    B_u &\equiv \begin{bmatrix}
        0_{5\times4} \\
        I_4
    \end{bmatrix} B_d \equiv \begin{bmatrix}
        0_{3\times3} & 0_{3\times3}\\
        I_3 & 0_{3\times3}\\
        0_{3\times3} & I_3
    \end{bmatrix}
\end{aligned}
\end{equation}
where $u_\zeta$ and $d_\nu$ represent the feedback control input and disturbance, respectively. The feedback control input $u_\zeta$ is designed by dynamic inversion, and thus the error dynamics in the Lie algebra can be written as:
\begin{equation}
\begin{aligned}
    \dot{\zeta} &= (-ad_{[\bar{\nu}_{v_r}]^\wedge} - C^\triangle + B_u K_\zeta) \zeta + B_d U_\zeta d_\nu \\
    u_\zeta &= U_\zeta^{-1}K_\zeta \zeta
\label{eq:zeta}
\end{aligned}
\end{equation}
where $K_\zeta$ is the control gain matrix for the state feedback controller, which can be designed using a Linear Quadratic Regulator (LQR) in the Lie algebra.

\subsection{Error Dynamics of Angular Velocity}
Consider the dynamics of the angular velocity of two systems, where $\omega^{eb}_b \in \mathbb{R}^3$ represents the angular velocity of the vehicle in the vehicle body frame. The angular velocity is relative to the earth frame, which is denoted by $^{eb}$. $\bar{\omega}^{er}_r$ represents the angular velocity of the reference trajectory in the reference frame.

We consider the error in the reference frame is $\omega^{br}_r = {\bar{\omega}}^{er}_r - R^{rb}{\omega}^{eb}_b$. The error dynamics can be written as:
\begin{equation}
    \dot{\omega}^{br}_r = \dot{\bar{\omega}}^{er}_r - R^{rb}\dot{\omega}^{eb}_b + \omega^{br}_r \times \bar{\omega}^{er}_r
\end{equation}
where $R^{rb}$ is the rotational matrix for rotating the vector from the vehicle body frame to the reference frame.

In our control design, we employ the dynamic inversion approach similar to~\cite{mellinger2011minimum}; therefore, our control law is:

\begin{equation}
\begin{aligned}
    \dot{\omega}^{eb}_b &= {R^{rb}}^{-1} (\dot{\bar{\omega}}^{er}_r - K_\omega {\omega}^{br}_r - d_\alpha) \\
    &= R^{br}(\dot{\bar{\omega}}^{er}_r - K_\omega {\omega}^{br}_r - d_\alpha)   
\end{aligned}
\end{equation}
where $K_\omega$ is the control gain matrix and $d_\alpha$ is the disturbance. The actual control in the multi-rotor, which is the moment of the motor in the vehicle body frame, is given by:
\begin{equation}
    M_b = \prescript{e}{}{}\frac{d}{dt} J_b \omega^{eb}_b = J \dot{\omega}^{eb}_b + \omega^{eb}_b \times J_b \omega^{eb}_b
\end{equation}
where $J_b$ is the inertia matrix of the vehicle in the vehicle body frame. Substituting the designed control law to the actual control, the moment in the body frame can be written as:
\begin{equation}  
    M_b = J_b R^{br} (\dot{\bar{\omega}}^{er}_r + K_\omega \omega^{br}_r - d_\alpha) + \omega^{eb}_b \times J_b \omega^{eb}_b
\end{equation}

% given bounded disturbance d_alpha
% assume bound on error_R
% calculate bound on e_omega from LMI
% calculate bound on R
% repeat until converged

The error dynamics can be rewritten as:
\begin{equation}
    \dot{\omega}^{br}_r = - {\bar{\omega}}^{er}_r \times {\omega}^{br}_r + K_\omega {\omega}^{br}_r + d_\alpha
\label{eq:e_omega}
\end{equation}
which is a linear system with bounded input.

\section{Application of Invariant Set}
\label{sec:IV}
Since the log-linear error dynamical system in \eqref{eq:zeta} is a linear system with bounded input, the invariant set can be found efficiently employing LMIs. To find the invariant set for the error bound of position, velocity, and rotation, we first apply LMIs to the error dynamics of angular velocity in \eqref{eq:e_omega} to find the error bound of the angular velocity, since the error dynamics in \eqref{eq:zeta} depends on the error of $\omega$. We then use the bound of the angular velocity as the bounded input for the error dynamics in \eqref{eq:zeta} and apply LMIs to find the invariant set for the log-linear system. Once we have the invariant set for the log-linear system in the $\mathfrak{se}_2(3)$ Lie algebra, we can compute the invariant set for the actual nonlinear system, which is the system in the $SE_2(3)$ Lie group, by applying the exponential map. Our LMI guarantees boundedness of the invariant set of the given system subject to bounded disturbance, but neglects sensor noise and estimator dynamics.

We consider a multi-rotor flying along a reference trajectory, which is generated by polynomial trajectory planning~\cite{mellinger2011minimum}; however, our algorithm is not limited to this case. The multi-rotor model given in \eqref{eq:sys} can be embedded in the $SE_2(3)$ Lie group, \eqref{eq:dX}. As we mentioned above, we first compute the feedback control with LQR and a Lyapunov function of the error dynamics of the angular velocity using an LMI in \eqref{eq:e_omega} with a reference angular velocity $[\omega_1, \omega_2, \omega_3] =  [5, 5, 1]\;rad/s$. We then use the Lyapunov function to compute the error bound of the angular velocity with a bounded external disturbance in $\alpha$, $||d_\alpha||_\infty = 0.1 \; rad/s^2$. The result of the maximum tracking error in the angular velocity is $||\omega||_\infty = 0.03162 \; rad/s$. 

% add the vector of acceleration (thrust vector in body frame), or sqaure to show the multi-rotor leaning along the reference trajectory
% \begin{figure}[tbp]
%     \centering
%     \includegraphics[width=0.8\columnwidth]{images/ref_traj.eps}
%     \caption{Reference Trajectory}
%     \label{fig:ref_traj}
% \end{figure}

\begin{figure}[htbp]
    \centering
    \begin{subfigure}[b]{0.49\textwidth}
        \includegraphics[width=\columnwidth]{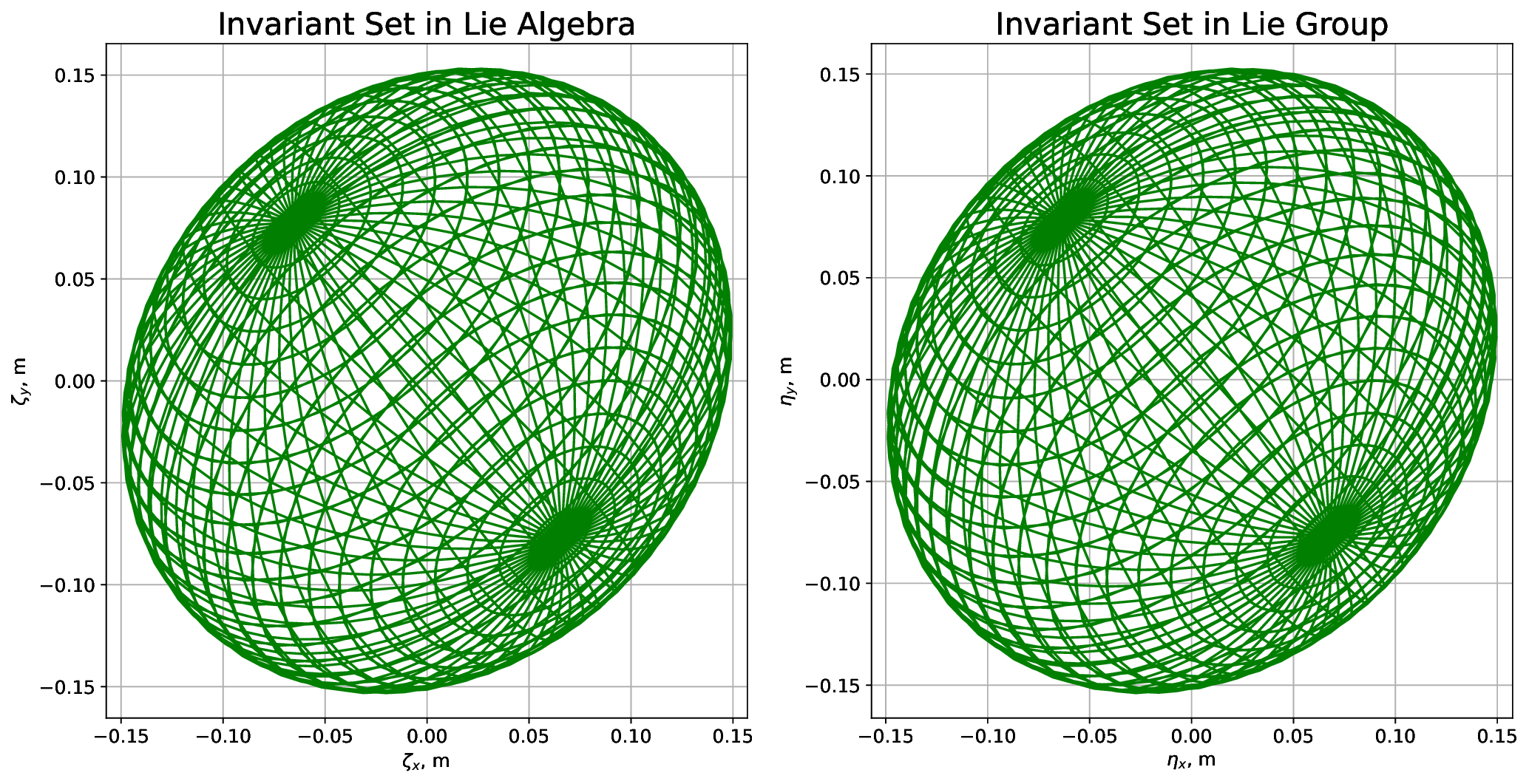}
    \end{subfigure}
    \begin{subfigure}[b]{0.49\textwidth}
        \includegraphics[width=\columnwidth]{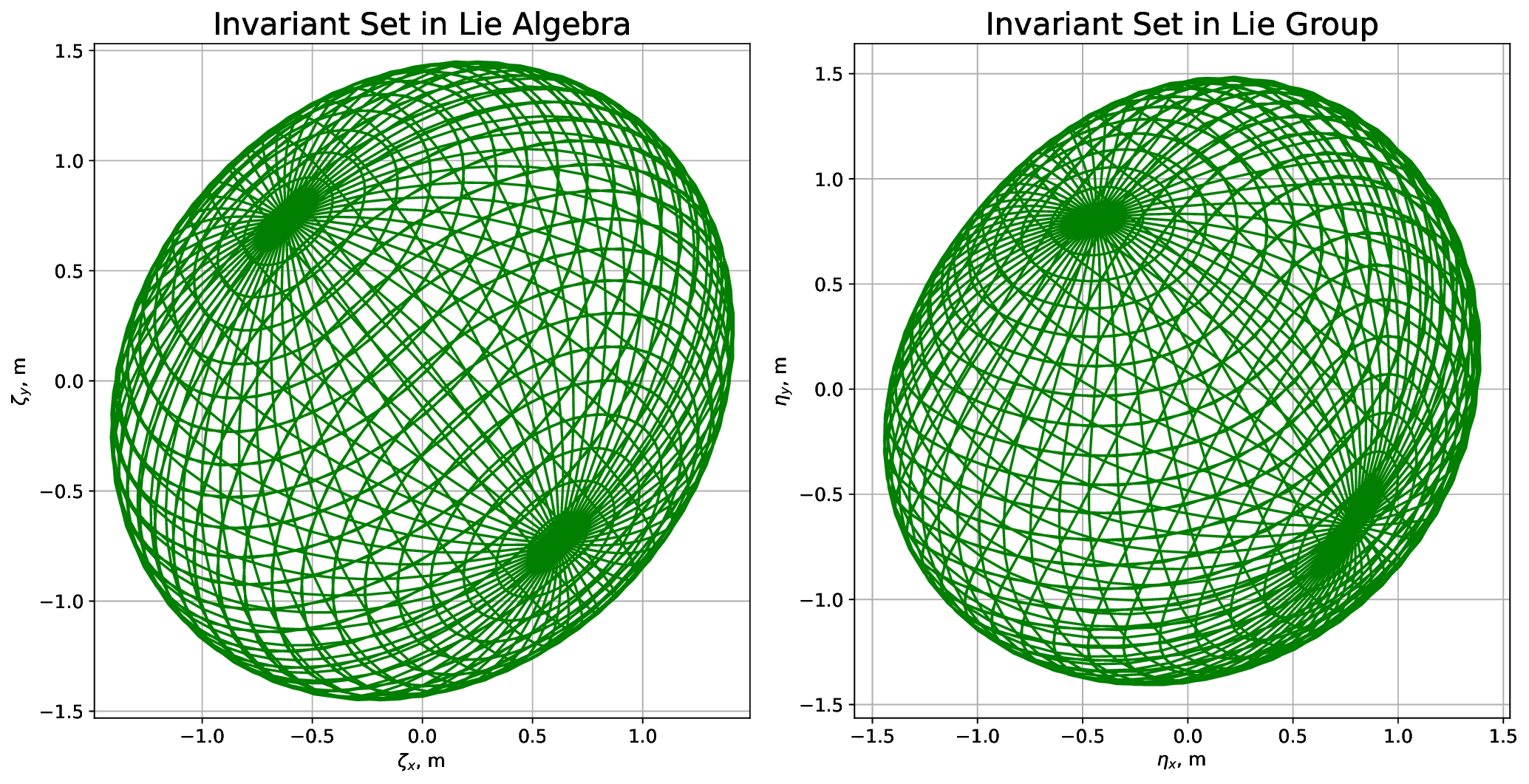}
    \end{subfigure}
    \caption{Projected 2D Invariant Sets Comparison with Dynamic Inversion, Left: Invariant Set in the Lie Algebra, Right: Invariant Sets in the Lie Group, Top: Small Disturbance, Bottom: Large Disturbance}
    \label{fig:invariant}
\end{figure}

\begin{figure}[htbp]
    \centering
    \begin{subfigure}[b]{0.5\textwidth}
        \includegraphics[width=\columnwidth]{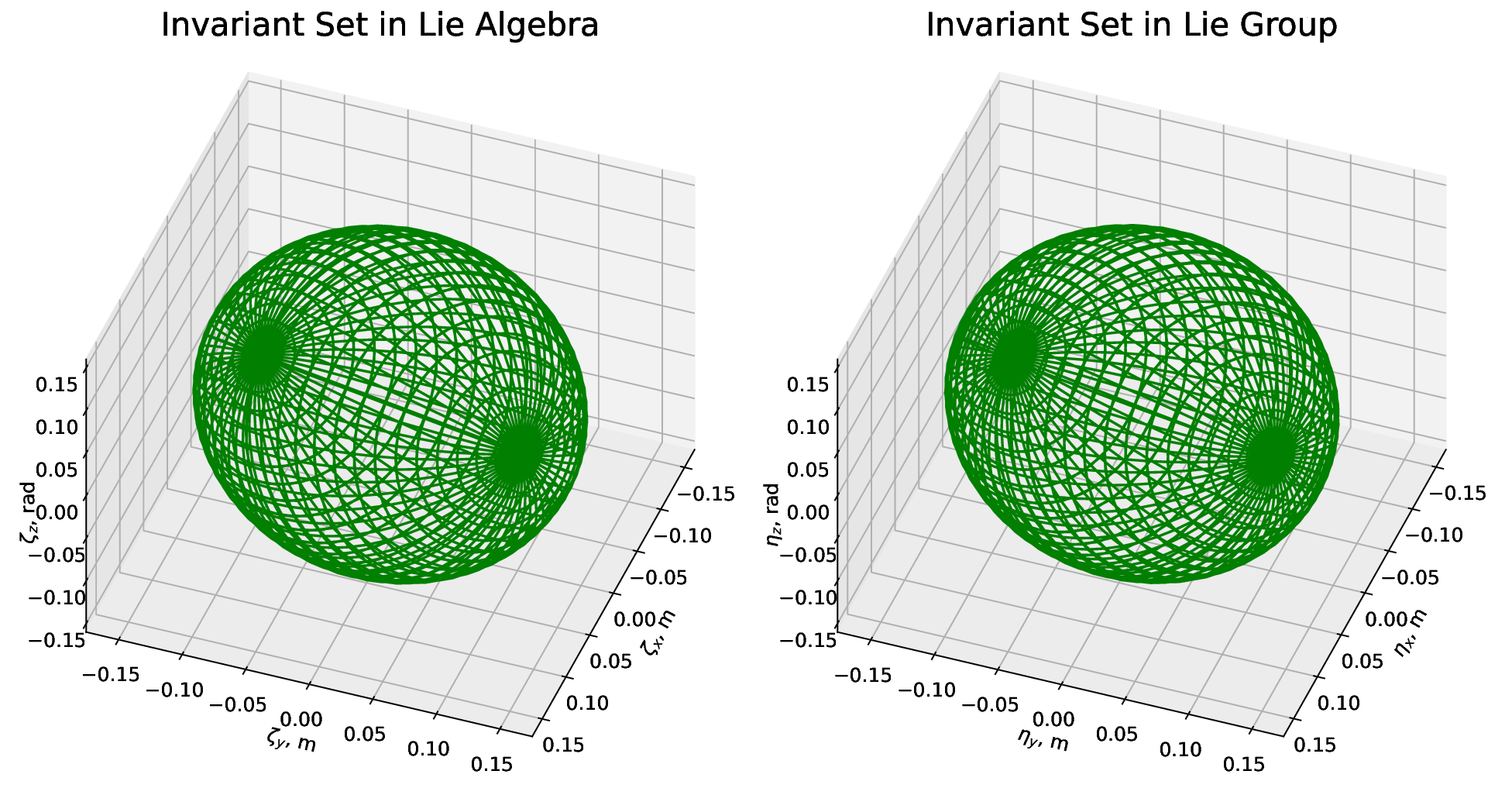}
    \end{subfigure}
    \begin{subfigure}[b]{0.5\textwidth}
        \includegraphics[width=\columnwidth]{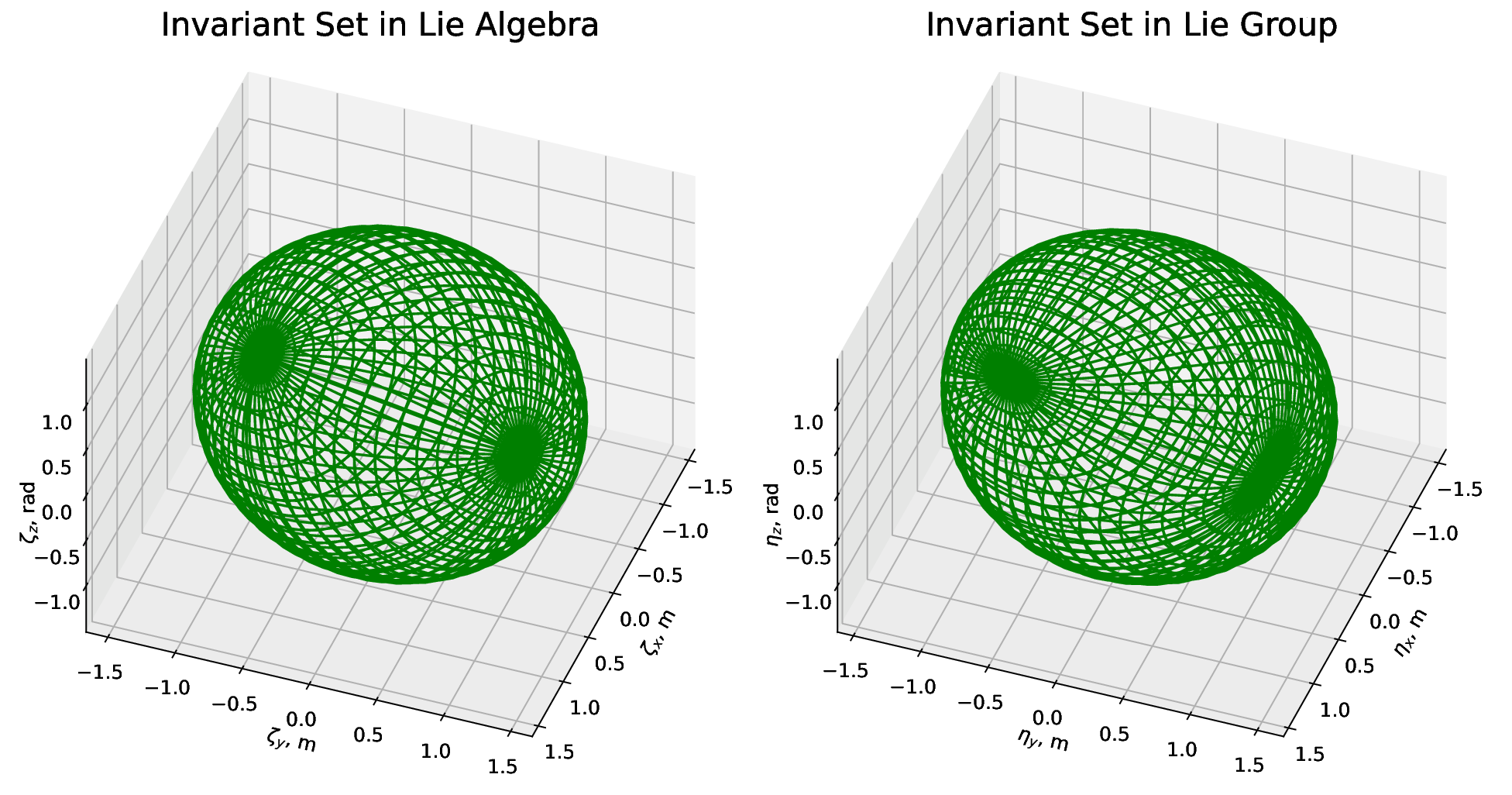}
    \end{subfigure}
    \caption{Projected 3D Invariant Sets Comparison with Dynamic Inversion, Left: Invariant Set in the Lie Algebra, Right: Invariant Sets in the Lie Group, Top: Small Disturbance, Bottom: Large Disturbance}
    \label{fig:invariant3d}
\end{figure}

The invariant set of the position error is computed with $[a_x, a_y, a_z] = [7.5, 7.5, 0]\;m/s^2$. For external disturbance in translational acceleration, we consider two different magnitudes. We assume the magnitude of bounded disturbance inputs, $||a||_\infty$, as $0.1\;m/s^2$ and $1\;m/s^2$. We use the LQR method to find the feedback controller for the log-linearized system in the $\mathfrak{se}_2(3)$ Lie algebra, and find the Lyapunov function using the LMIs. The final result of the invariant sets in the Lie algebra and the Lie group is shown in \Cref{fig:invariant} and \Cref{fig:invariant3d}, respectively. The figures show the invariant set with zero initial states in the $SE_2(3)$ Lie group, and the invariant set in the Lie group is constructed by applying the exponential map from the set in the Lie algebra. The Lyapunov function obtained in the LMI approach is an ellipsoid, which is the shape of the invariant sets for the log-linearized system in the $\mathfrak{se}_2(3)$ Lie algebra as shown on the left of \Cref{fig:invariant}.

\begin{figure}[tbp]
    \centering
    \begin{subfigure}[t]{0.5\textwidth}
    \centering
        \includegraphics[width=0.498\columnwidth]{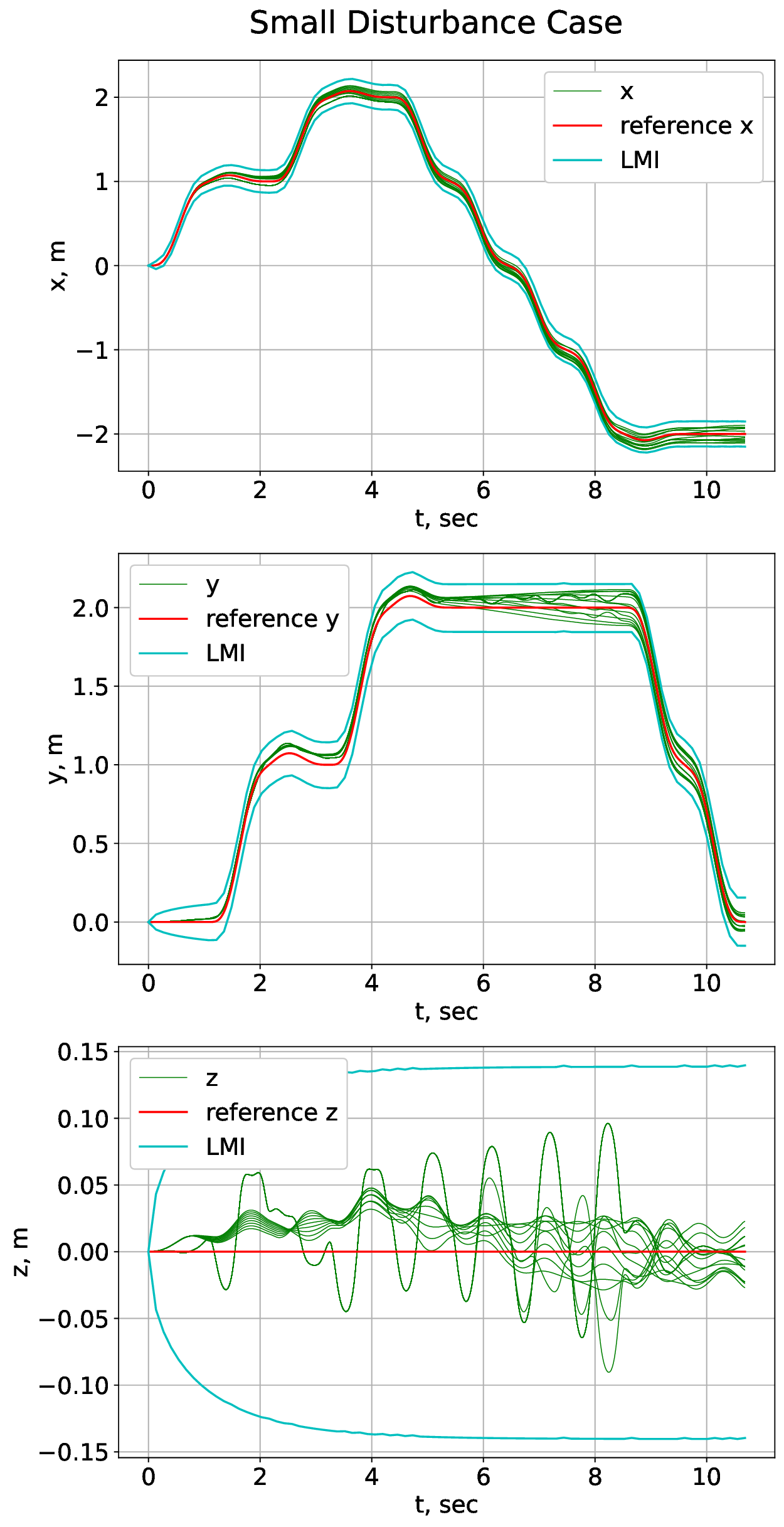}
        \includegraphics[width=0.49\columnwidth]{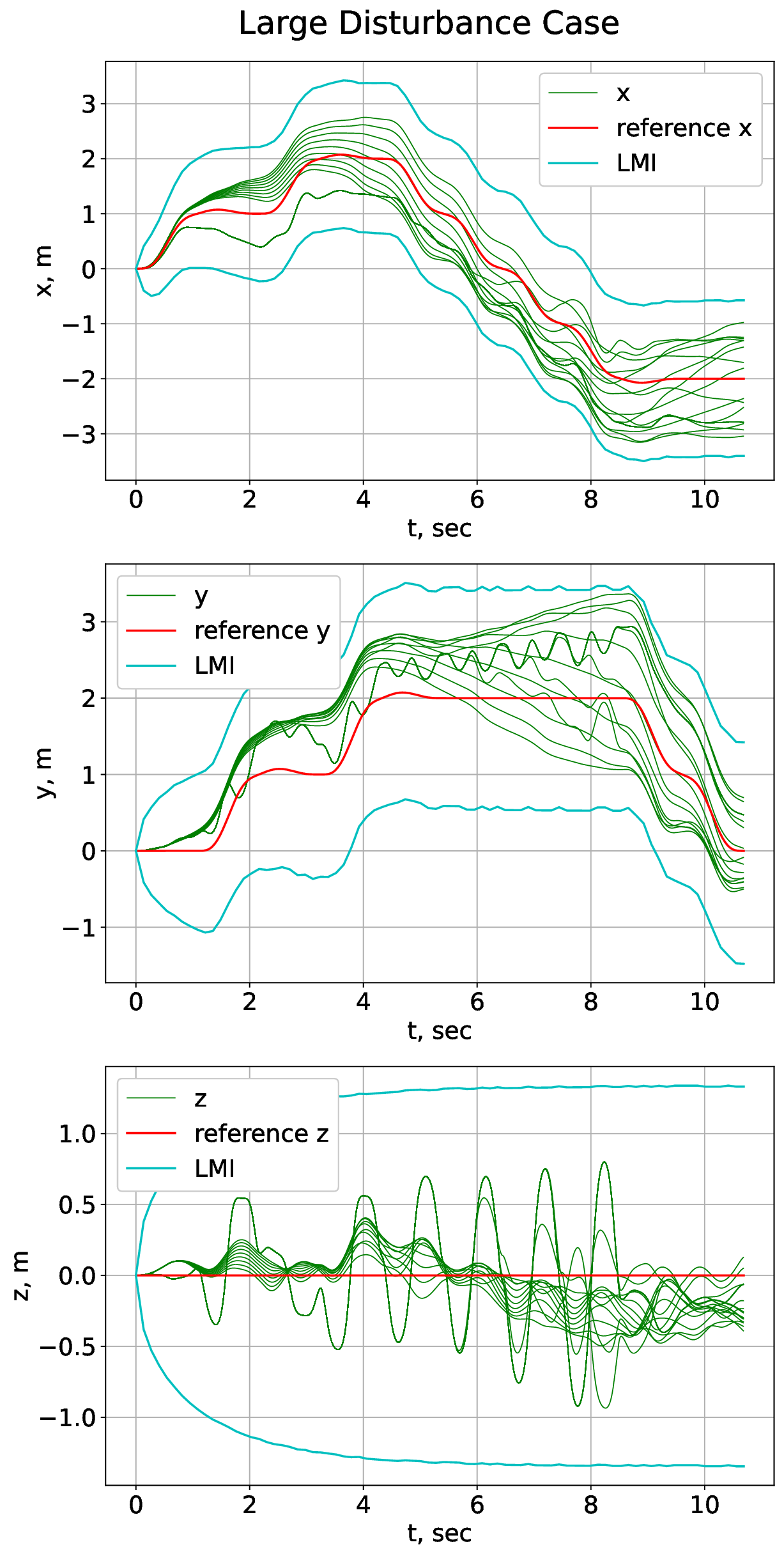}
    \end{subfigure}
    \caption{The Bound of the Invariant Set with Simulated Trajectories, Left: Small Disturbance, Right: Large Disturbance}
    \label{fig:invariant_bound}
\end{figure}

Figure~\ref{fig:invariant_bound} shows a time history plot of the invariant set bound computed by our approach, along with simulated trajectories using our dynamic inversion controller and both sinusoidal and square wave disturbances randomly sample. The left figure in~\Cref{fig:invariant_bound} shows the scenario under small disturbance, while the right figure shows the scenario under large wind disturbance. In both cases, the invariant sets calculated from the LMIs successfully bound all simulated trajectories. The precision of the invariant set shown in~\Cref{fig:invariant_bound} can be improved by doing an iterative process between our log-linearized system in~\Cref{eq:zeta} and sub-system in~\Cref{eq:e_omega}. This is a useful result, since it can be applied to safety verification of multi-rotors with a more complicated reference trajectory.

\section{Conclusion}
\label{sec:V}
In this paper, we presented an efficient method to compute the invariant sets for the error dynamics of a multi-rotor with the proposed log-linear dynamics inversion controller with bounded disturbances. Our method was based on log-linearization in the Lie group, which allowed us to exactly log-linearize the nonlinear system. Additionally, we utilized the Linear Matrix Inequalities (LMIs) for a linear system with bounded inputs to bound the tracking error. Our simulation demonstrated the application of our algorithm to a multi-rotor with a reference trajectory and showed the usefulness of our algorithm. 

As future work, we intend to consider the sensor noise and estimator dynamics in our multi-rotor model to have a more precise computation, and apply our approach with flow pipe creation with a more complicated reference trajectory in order to illustrate a more realistic Urban Air Mobility (UAM) scenario. 

\bibliographystyle{IEEEtran}
\bibliography{ref}

\end{document}